\documentclass[12pt]{article}
\usepackage{amsfonts}
\usepackage{latexsym}
\usepackage{amsmath,amssymb}
\usepackage{verbatim}
\usepackage{setspace}
\usepackage{color}
\usepackage{physics}
\usepackage{tikz}
\usepackage{mathdots}
\usepackage{yhmath}
\usepackage{cancel}
\usepackage{color}
\usepackage{siunitx}
\usepackage{mathrsfs}  
\usepackage{array}
\usepackage{multirow}
\usepackage{amssymb}
\usepackage{gensymb}
\usepackage{tabularx}
\usepackage{hyperref}
\usepackage{booktabs}
\usetikzlibrary{fadings}
\usetikzlibrary{patterns}
\usetikzlibrary{shadows.blur}
\usetikzlibrary{shapes}

\usepackage[textheight=9in, textwidth=6.5in, letterpaper]{geometry}

\numberwithin{equation}{section}

\def\ip{${\mathcal I}^+$}

 \def\p{\partial}
 \newcommand{\Oc}{\mathcal{O}}
 \def\bz{{\bar z}}
 
\def\0{{(0)}}
\def\1{{(1)}}
\def\2{{(2)}}

\def\n{\nabla}

\def\<{\langle }
\def\>{\rangle }
\def\[{\left[}
\def\]{\right]}
\def\bw{{\bar w}}

\newcommand{\bea}{\begin{eqnarray}}
\newcommand{\eea}{\end{eqnarray}}
\newcommand{\be}{\begin{equation}}
\newcommand{\ee}{\end{equation}}
\newcommand{\ba}{\begin{align}}
\newcommand{\ea}{\end{align}}

\renewcommand{\epsilon}{\varepsilon}

   \makeatletter
  \let\over=\@@over \let\overwithdelims=\@@overwithdelims
  \let\atop=\@@atop \let\atopwithdelims=\@@atopwithdelims
  \let\above=\@@above \let\abovewithdelims=\@@abovewithdelims
\renewcommand\section{\@startsection {section}{1}{\z@}%
                                   {-3.5ex \@plus -1ex \@minus -.2ex}
                                   {2.3ex \@plus.2ex}%
                                   {\normalfont\large\bfseries}}

\renewcommand\subsection{\@startsection{subsection}{2}{\z@}%
                                     {-3.25ex\@plus -1ex \@minus -.2ex}%
                                     {1.5ex \@plus .2ex}%
                                     {\normalfont\bfseries}}

\begin{document}
\unitlength = 1mm
\ \\
\vskip1cm
\begin{center}

{ \LARGE {\textsc{Celestial Feynman Rules for Scalars}}}

\vspace{0.8cm}
Walker Melton

\vspace{1cm}

{\it  Center for the Fundamental Laws of Nature, Harvard University,\\
Cambridge, MA 02138, USA}

\begin{abstract}
Off-shell celestial amplitudes with both time-like and space-like external legs are defined. The Feynman rules for scalar amplitudes, viewed as a set of recursion relations for off-shell momentum space amplitudes, are transformed to the celestial sphere using the split representation. For four-point celestial amplitudes, the Feynman expansion is shown to be equivalent to a conformal partial wave decomposition, providing an interpretation of  conformal partial wave expansion coefficients as integrals over off-shell three-point structures. A conformal partial wave decomposition for a simple four-point $s$-channel massless scalar celestial amplitude is derived.
 \end{abstract}
\vspace{0.5cm}

\vspace{1.0cm}

\end{center}

\pagestyle{empty}
\pagestyle{plain}
\newpage
\tableofcontents
\def\gzz{\gamma_{z\bz}}
\def\vx{{\vec x}}
\def\p{\partial}
\def\po{$\cal P_O$}
\def\cN{{\cal N}_\rho^2 }
\def\N{${\cal N}_\rho^2 ~~$}
\def\G{\Gamma}
\def\a{{\alpha}}
\def\b{{\beta}}
\def\g{\gamma}
\def\ch{{\cal H}^+}
\def\hf{{\cal H}}
\def\hbh{{\cal H}_{\rm BH}}
\def\hout{{\cal H}_{\rm OUT}}
\def\ss{\Sigma_S}
\def\D{{\rm \Delta}}
\def \bw {{\bar w}}
\def \bz {{\bar z}}
\def\cS{{\cal S}}
\def\l{\lambda }
\def\d{{\delta}}
\def\n{{\rm SC}}
\def\ip{{\rm cft}}
\def\RR{\mathbb{K}}
\def\i{i^\prime}
\def\adz{AdS$_3/\mathbb{Z}$}
\def\sll{$SL(2, \mathbb{R})_L$}
\def\slr{$SL(2, \mathbb{R})_R$}
\pagenumbering{arabic}

\section{Introduction}

The AdS/CFT correspondence, which relates theories of quantum gravity in asymptotically anti-de-Sitter spacetimes to lower-dimensional conformal field theories, has been one of the key tools for understanding quantum gravity. It is therefore important to understand if this correspondence can be generalized to more physically realistic asymptotically flat and asymptotically de-Sitter spacetimes \cite{jdbs}. Celestial holography has recently been proposed as a hypothetical 2D CFT dual to flat-space quantum gravity. Scattering amplitudes between states that diagonalize Lorentz transformations, rather than translations, take the form of correlation functions of primaries in a 2D theory with a global conformal invariance \cite{fsa2017, gluonamp}. When coupled to gravity, the global conformal symmetry is promoted to a local conformal symmetry and a stress tensor can be defined as a shadow of the soft graviton \cite{stresstensor}. Thus, there has been intense interest in understanding if the conformal field theory defined by identifying scattering amplitudes between boost eigenstates with correlation functions is well defined, if it provides a suitable holographic dual to flat-space scattering, and how techniques developed to study CFTs may be applied to the scattering problem in quantum gravity. 

Much progress has been made. Celestial soft theorems for gauge and gravitational theories have been derived and shown to imply the presence of an infinite tower of currents \cite{confsoft, confsoftgrav, holsymalg}. Operator product expansions have been derived from colinear singularities of momentum-space amplitudes  \cite{celOPE,bmsope}. The imprints of bulk translation and BMS symmetries on celestial amplitudes have been understood \cite{poinconst,BMSI,SBMSI}. Celestial holography has been extended to scattering in (2,2)-signature Klein space \cite{klein}. The  state-operator correspondence in celestial CFTs has been defined \cite{stateop}. Nevertheless, basic questions remain: namely, how are higher-point celestial amplitudes built from lower-point amplitudes? 

In momentum space, higher-point scattering amplitudes are well-understood to derive from lower point amplitudes, both through off-shell Feynman expansions and various on-shell recursion relations, such as the Britto-Cachazo-Feng-Witten recursion relation in Yang-Mills theory \cite{BCFW}. In conformal field theories, higher-point functions are understood to arise from a basic set of two- and three-point correlators through the operator product expansion \cite{df}. Understanding how higher-point celestial amplitudes are built up from lower-point celestial amplitudes, and whether there exists a `celestial OPE,' are important steps in understanding celestial CFT as a hypothetical conformal dual to flat space quantum gravity. Fan, Fotopoulos, Stieberger, Taylor, and Zhu found a conformal block decomposition of four-point gluon amplitudes after shadowing an external leg \cite{fan2021}. Lam and Shao showed that the optical theorem for celestial amplitudes takes the form of a conformal partial wave decomposition of the imaginary part of the amplitude \cite{shaolam}. Atanasov, Melton, Raclariu, and Strominger derived a conformal block decomposition for scalar celestial amplitudes in (2,2)-signature Klein space where the conformal block coefficients took the form of celestial three-point coefficients up to simple trigonometric factors \cite{CPW}. In a CFT, the three-point structures are sufficient to reconstruct four- and higher-point functions, so we expect that in a celestial CFT one should be able to reconstruct arbitrary celestial amplitudes, including the real part, from three-point data. Yet most work either reconstructs solely the imaginary part of the amplitude, works backwards from a known four-point function found by transforming momentum-space amplitudes to the conformal primary basis, or directly transforms known soft and colinear limits to the celestial sphere.

In this work, we demonstrate that one can transform momentum-space Feynman rules, viewed as recursion relations for off-shell momentum space amplitudes, to the celestial sphere. This provides a method to reconstruct higher-point celestial amplitudes from lower-point amplitudes, provided that one can analytically continue them off-shell. We first define off-shell conformal primary wavefunctions for both positive and negative $-p^2 = M^2$ in section \ref{oscpw}. In section \ref{osca}, we define off-shell celestial amplitudes with both time-like and space-like external legs, which appear in the celestial Feynman rules derived in section \ref{cfr}. In section \ref{fpf}, we specialize the relationship to four-point tree-level functions and show that it takes the form of a conformal partial wave expansion with coefficients given by integrals over products of off-shell three-point structures. The implications of this expansion to factorization in the celestial theory are discussed. In section \ref{scalarfpf}, we use this relationship to derive conformal partial wave expansions for a simple scalar celestial amplitude. Section \ref{loops} concludes with comments on loop integrals.

\section{Preliminaries}

Celestial CFT investigates scattering  between boost eigenstates, recasting amplitudes as $n$-point functions on the celestial sphere that transform as correlation functions of highest-weight operators under the Lorentz SL(2,$\mathbb{C}$) symmetry \cite{fsa2017,confbas}. In this section, we review the formalism for celestial amplitudes of massless and massive scalars. 

 Null momenta are parametrized by 
\begin{equation}
q^\mu(z,\bar{z}) = \omega\hat{q}^\mu(z,\bar{z})
\end{equation}  
where $\hat{q}^\mu$ is the unit null momenta 
\begin{equation}
\hat{q}^\mu(z,\bar{z}) = (1+z\bar{z},z-\bar{z},-i(z-\bar{z}),1-z\bar{z})
\end{equation}
 and $z = \bar{z}^* \in \mathbb{C}$. Massive time-like unit momenta $\hat{p}^2 = -1$ are parametrized by 
\begin{equation}
\hat{p}^\mu(y,z,\bar{z}) = \frac{1}{2y}(1+y^2+z\bar{z},z+\bar{z},-i(z-\bar{z}),1-y^2-z\bar{z}).
\end{equation}

Celestial amplitudes describe scattering between particles in conformal primary wavefunctions, an alternative basis for solutions of the on-shell wave equation that form highest-weight representations of the Lorentz group \cite{confbas,liulowe}. These transform as 
\begin{equation}
    \phi_{(\Delta,M^2)}^\pm \left(\Lambda^\mu_{\ \nu}X^\nu\Big|\frac{az+b}{cz+d},\frac{\bar{a}\bar{z}+\bar{b}}{\bar{c}\bar{z}+\bar{d}}\right) = |cw+d|^{2\Delta}\phi_{(\Delta,M^2)}^\pm(X|z,\bar{z})
\end{equation}
under Lorentz transformations.

For a massive scalar, these wavefunctions take the form 
\begin{equation}
    \phi^\pm_\Delta(X|z,\bar{z}) = \int_0^\infty\frac{dy}{y^3}\int dwd\bar{w}G_{\Delta}(y,w,\bar{w}|z,\bar{z})e^{\pm im\hat{p}(y,w,\bar{w}) \cdot X}
\end{equation}
where $G_\Delta$ is the H$_3$ bulk-to-boundary propagator 
\begin{equation}
G_\Delta(y,w,\bar{w}|z,\bar{z}) = \left(\frac{y}{y^2+|z-w|^2}\right)^\Delta.
\end{equation}
These form a $\delta$-function normalizable basis for solutions to the wave equation for $z,\bar{z} \in \mathbb{C}$ and $\Delta \in 1+i\mathbb{R}_+$. The measure is related to the Lorentz invariant measure $\widetilde{d^3\vec{p}}$ by 
\begin{equation}
\widetilde{d^3\hat{p}}= \frac{d^3\hat{p}}{2\hat{p^0}} = \frac{1}{2}\frac{dydzd\bar{z}}{y^3}.
\end{equation}
For a massless particle the conformal primary wavefunctions take the form 
\begin{equation}
    \phi^\pm_\Delta(X|z,\bar{z}) = \int_0^\infty d\omega\omega^{\Delta-1}e^{\pm i\omega\hat{q}(z,\bar{z}) \cdot X}.
\end{equation}
These form a $\delta$-function normalizable basis for solutions to the massless Klein-Gordon equation for $z,\bar{z} \in \mathbb{C}$ and $\Delta =1 + i\mathbb{R}$ \cite{confbas}.

Celestial amplitudes are labelled by a boost weight $\Delta$ and a point $(z,\bar{z})$ on the celestial sphere for each external particle. If the momentum space amplitude, including the momentum-conserving $\delta$-function, is $A(p_1,\ldots,p_m,q_{m+1},q_n)$ where $p_i$ label particles of mass $m_i$ and $q_j$ label massless particles, the celestial amplitude is 
\begin{equation}
\begin{split}
\tilde{A}_{\Delta_1,\ldots,\Delta_n}^{\epsilon_1,\ldots,\epsilon_n}(z_1,\bar{z}_1,\ldots,z_n,\bar{z}_n) &= \int \prod_{i=1}^m \frac{dy_idw_id\bar{w}_i}{y_i^3}G_{\Delta_i}(y_i,w_i,\bar{w}_i|z_i,\bar{z}_i) \prod_{j=m+1}^n d\omega_i\omega_i^{\Delta_i-1} \\
&\times A(\epsilon_1m_1\hat{p}(y_1,w_1,\bar{w}_1),\ldots,\epsilon_{m+1}\omega_{m+1}q(z_{m+1},\bar{z}_{m+1}),\ldots,\epsilon_n\omega_nq(z_n,\bar{z}_n))	
\end{split}
\end{equation}
where $\epsilon_i = -1$ ($1$) if particle $i$ is ingoing (outgoing). 

\section{Off-Shell Conformal Primary Wavefunctions}
\label{oscpw}
An off-shell conformal primary wavefunction $\phi^\pm_{(\Delta,M^2)}$ is a solution to the wave equation
\begin{equation}
    (\partial^2-M^2)\phi^\pm_{(\Delta,M^2)} = 0
\end{equation}
that transforms under Lorentz transformations as a conformal primary of weight $\Delta$, so that under Lorentz transformations
\begin{equation}
    \phi_{(\Delta,M^2)}^\pm \left(\Lambda^\mu_{\ \nu}X^\nu\Big|\frac{az+b}{cz+d},\frac{\bar{a}\bar{z}+\bar{b}}{\bar{c}\bar{z}+\bar{d}}\right) = |cw+d|^{2\Delta}\phi_{(\Delta,M^2)}^\pm(X|z,\bar{z}).
\end{equation}
These off-shell conformal primaries are examples of generalized conformal primaries that solve the Klein-Gordon equation of mass $M \in \mathbb{C}$ rather than the equation of motion of the field \cite{shifting}. They form a basis for normalizable wavefunctions that may or may not satisfy the equations of motion.
\subsection{Off-Shell Conformal Primary Wavefunctions for $M^2 > 0$}
For $M^2 > 0$, the off-shell conformal primary wavefunction takes the form 
\begin{equation}
\label{pmcpw}
    \phi^\pm_{(\Delta,M^2)}(X|z,\bar{z}) = \int_0^\infty\frac{dy}{y^3}\int dwd\bar{w}G_\Delta(y,w,\bar{w}|z,\bar{z})e^{\pm iM\hat{p}\cdot X}.
\end{equation}
Evaluating this integral for $M \in -i\mathbb{R}_+$ and analytically continuing gives \cite{fsa2017}
\begin{equation}
    \phi^\pm_{(\Delta,M^2)}(X|z,\bar{z}) = \frac{4\pi}{iM}\frac{(\sqrt{-X^2})^{\Delta-1}}{(-X^\mu q_\mu \mp i\epsilon)^\Delta}K_{\Delta-1}(iM\sqrt{-X^2}).
\end{equation}

\subsection{Off-Shell Conformal Primary Wavefunctions for $M^2 < 0$}
To define celestial amplitudes with spacelike external legs ($-M^2 = p^2 = \mu^2 > 0)$, we adapt the technology developed for time-like external legs. A generic unit spacelike momentum  $\hat{p}_+^2 = 1$ can be parametrized as 
\begin{equation}
\hat{p}_+ = \frac{1}{2\eta}\left(1-\eta^2+z\bar{z},z+\bar{z},-i(z-\bar{z}),1+\eta^2-z\bar{z}\right).
\end{equation} 
In these coordinates, the induced metric takes the form 
\begin{equation}
ds^2_{dS} = \frac{-d\eta^2 + dzd\bar{z}}{\eta^2}.
\end{equation}

To ensure that celestial amplitudes with spacelike external legs are consistent with analytic continuation from $M^2 > 0$, we define the bulk-to-boundary propagator for the dS$_3$ slice by 
\begin{equation}
G_\Delta^{dS}(\eta,z,\bar{z}) = -G_\Delta(-i\eta,z,\bar{z}).
\end{equation}
We derive this by analytic continuation from $M^2 > 0$ in Appendix \ref{ancon}.

\begin{figure}[h]
\begin{center}
\includegraphics[width=4in]{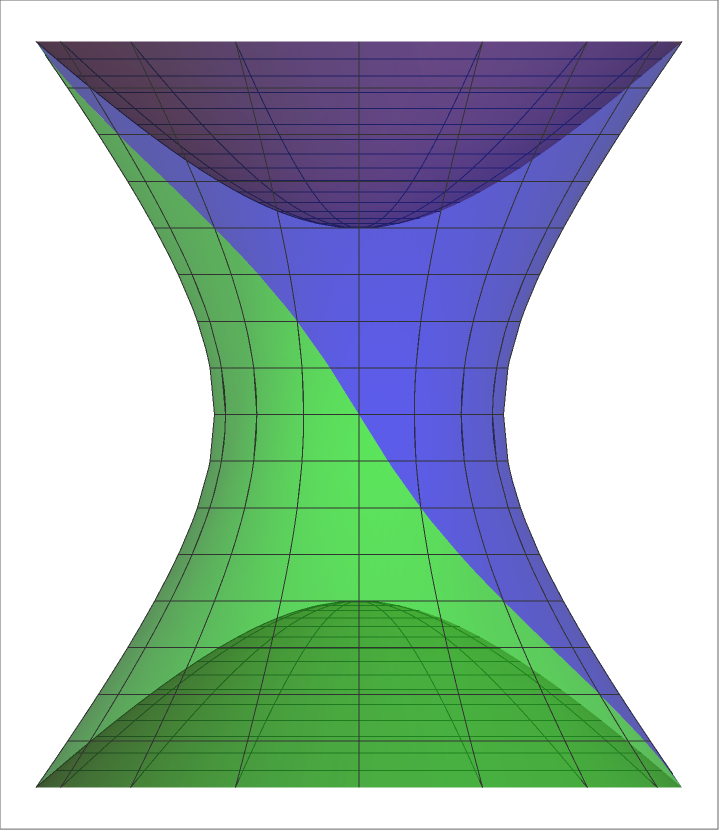}
\end{center}
\caption{Integration contours for the integral transform to the celestial primary wavefunctions for massive outgoing timelike (purple), massive ingoing timelike (dark green), spacelike outgoing (blue) and spacelike ingoing (green) external momenta.}
\end{figure}

The off-shell celestial amplitudes for $M^2 < 0$ therefore take the form 
\begin{equation}
\label{osnmp}
\begin{split}
    \phi^\pm_{(\Delta,-\mu^2)}(X|z,\bar{z}) &= \int_0^\infty \frac{d\eta}{\eta^3}\int dwd\bar{w}G_\Delta^{dS}(\eta,w,\bar{w}|z,\bar{z})e^{\pm i\mu \hat{p}_+ \cdot X} \\
    &= \frac{4\pi}{\mu}\frac{(\sqrt{-X^2})^{\Delta-1}}{(-X^\mu q_\mu\pm i\epsilon)^\Delta}K_{\Delta-1}(\mu\sqrt{-X^2}).
\end{split}
\end{equation}
The choice of $G^{dS}_\Delta$ guarantees that the wavefunction in Equation \ref{osnmp} agrees with analytic continuation $M \to -i\mu$. We choose the convention that outgoing  solutions are integrated over the region $\eta > 0$ as this region contains the future celestial sphere. 
\section{Off-Shell Celestial Amplitudes}
\label{osca}
Feynman rules provide a convenient recursion relation for off-shell momentum space amplitudes $A(p_i)$ for $p_i^2$ unconstrained. In this section, we define off-shell celestial amplitudes as hyperbolic transforms of off-shell momentum space amplitudes. These amplitudes depend on a conformal weight $\Delta$, a point on the celestial sphere $(z,\bar{z})$, and $M^2 = -p^2$, and are related to off-shell momentum-space amplitudes by the integral transforms described in Equations \ref{pmcpw} and \ref{osnmp}. Because taking $M^2 \to m^2$, where $m^2$ is the physical mass of the particle, recovers on-shell conformal primary wavefunctions, taking the same limit on the external legs recovers the on-shell celestial amplitudes commonly studied.
\subsection{Off-Shell Celestial Amplitudes for Time-Like External Legs} 
For external legs with timelike momenta (positive $M^2 = -p^2$), we define the off-shell conformal primary amplitudes

\begin{equation}
	\tilde{A}^{\epsilon_1\ldots\epsilon_n}_{(\Delta_1,M_1^2),\ldots,(\Delta_n,M_n^2)}(z_1,\ldots,z_n) = \int \prod_{i=1}^n \frac{dy_idw_id\bar{w}_i}{y_i^3}G_{\Delta_i}(y_i,w_i,\bar{w}_i|z_i,\bar{z}_i)A(\epsilon_i M_i\hat{p}(y_i,w_i,\bar{w}_i)).
\end{equation}
Taking $M_i$ to the on-shell mass $m_i$ then reproduces the on-shell celestial amplitude $A^{\epsilon_i}_{\Delta_i}(z_i,\bar{z}_i)$. 

\subsection{Off-Shell Celestial Amplitudes for Space-Like External Legs}

From the transformation from momentum-space to the off-shell conformal primary wavefunction defined in Equation \ref{osnmp}, we see that off-shell celestial amplitudes with spacelike ($p^2 = \mu^2, \mu > 0$) external legs can then be defined by 
\begin{equation}
    \label{ossla}
    	\tilde{A}^{\epsilon_1\cdots}_{(\Delta_1,-\mu_1^2)\cdots}(z_1,\bar{z}_1,\ldots) = \int_0^\infty\frac{d\eta}{\eta^3}\int dwd\bar{w} G^{dS}_{\Delta}(\eta,w,\bar{w}|z,\bar{z})A(\epsilon_1\mu_1\hat{p}^+(\eta,w,\bar{w}),\ldots)
\end{equation}
where legs $j=2,\ldots,n$ are transformed to the celestial sphere appropriately $-p_j^2$ greater than or less than 0. The choice of bulk-to-boundary propagator guarantees that Equation \ref{ossla} agrees with naive analytic continuation $M \to -i\mu$, as can be seen from Appendix \ref{ancon}. 
\section{Celestial Feynman Rules}
\label{cfr}
Consider the amplitude corresponding to the following one-particle-reducible diagram:
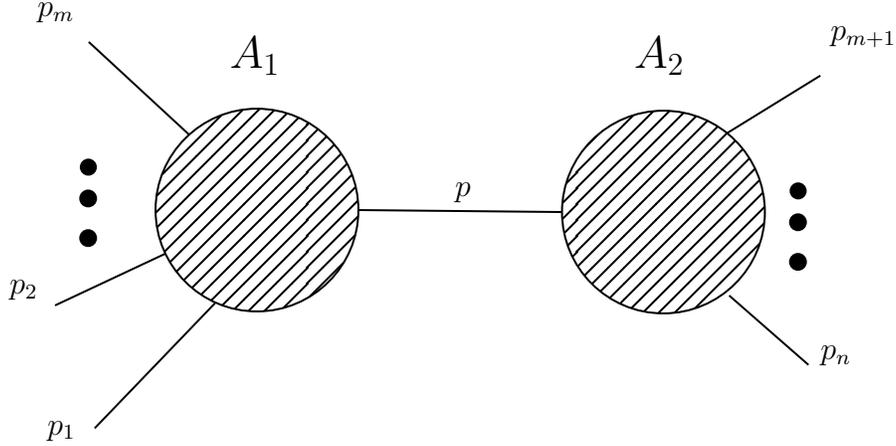
\begin{figure}[h!]
\begin{center}

 
\tikzset{
pattern size/.store in=\mcSize, 
pattern size = 5pt,
pattern thickness/.store in=\mcThickness, 
pattern thickness = 0.3pt,
pattern radius/.store in=\mcRadius, 
pattern radius = 1pt}
\makeatletter
\pgfutil@ifundefined{pgf@pattern@name@_93x1arahl}{
\pgfdeclarepatternformonly[\mcThickness,\mcSize]{_93x1arahl}
{\pgfqpoint{0pt}{0pt}}
{\pgfpoint{\mcSize+\mcThickness}{\mcSize+\mcThickness}}
{\pgfpoint{\mcSize}{\mcSize}}
{
\pgfsetcolor{\tikz@pattern@color}
\pgfsetlinewidth{\mcThickness}
\pgfpathmoveto{\pgfqpoint{0pt}{0pt}}
\pgfpathlineto{\pgfpoint{\mcSize+\mcThickness}{\mcSize+\mcThickness}}
\pgfusepath{stroke}
}}
\makeatother

 
\tikzset{
pattern size/.store in=\mcSize, 
pattern size = 5pt,
pattern thickness/.store in=\mcThickness, 
pattern thickness = 0.3pt,
pattern radius/.store in=\mcRadius, 
pattern radius = 1pt}
\makeatletter
\pgfutil@ifundefined{pgf@pattern@name@_ggw43h9iw}{
\pgfdeclarepatternformonly[\mcThickness,\mcSize]{_ggw43h9iw}
{\pgfqpoint{0pt}{0pt}}
{\pgfpoint{\mcSize+\mcThickness}{\mcSize+\mcThickness}}
{\pgfpoint{\mcSize}{\mcSize}}
{
\pgfsetcolor{\tikz@pattern@color}
\pgfsetlinewidth{\mcThickness}
\pgfpathmoveto{\pgfqpoint{0pt}{0pt}}
\pgfpathlineto{\pgfpoint{\mcSize+\mcThickness}{\mcSize+\mcThickness}}
\pgfusepath{stroke}
}}
\makeatother
\tikzset{every picture/.style={line width=0.75pt}} 

\begin{tikzpicture}[x=0.75pt,y=0.75pt,yscale=-1,xscale=1]

\draw  [pattern=_93x1arahl,pattern size=6pt,pattern thickness=0.75pt,pattern radius=0pt, pattern color={rgb, 255:red, 0; green, 0; blue, 0}] (100,179.15) .. controls (100,150.9) and (122.9,128) .. (151.15,128) .. controls (179.41,128) and (202.31,150.9) .. (202.31,179.15) .. controls (202.31,207.41) and (179.41,230.31) .. (151.15,230.31) .. controls (122.9,230.31) and (100,207.41) .. (100,179.15) -- cycle ;
\draw  [pattern=_ggw43h9iw,pattern size=6pt,pattern thickness=0.75pt,pattern radius=0pt, pattern color={rgb, 255:red, 0; green, 0; blue, 0}] (305,180.15) .. controls (305,151.9) and (327.9,129) .. (356.15,129) .. controls (384.41,129) and (407.31,151.9) .. (407.31,180.15) .. controls (407.31,208.41) and (384.41,231.31) .. (356.15,231.31) .. controls (327.9,231.31) and (305,208.41) .. (305,180.15) -- cycle ;
\draw    (202.31,179.15) -- (305,180.15) ;
\draw    (69.31,289.29) -- (130,226.15) ;
\draw    (49.31,227.29) -- (105,201.15) ;
\draw    (66.31,94.29) -- (117,141.15) ;
\draw  [fill={rgb, 255:red, 0; green, 0; blue, 0 }  ,fill opacity=1 ] (62.01,193.3) .. controls (62.01,191.1) and (63.8,189.31) .. (66.01,189.31) .. controls (68.21,189.31) and (70,191.1) .. (70,193.3) .. controls (70,195.51) and (68.21,197.29) .. (66.01,197.29) .. controls (63.8,197.29) and (62.01,195.51) .. (62.01,193.3) -- cycle ;
\draw  [fill={rgb, 255:red, 0; green, 0; blue, 0 }  ,fill opacity=1 ] (62.01,173.3) .. controls (62.01,171.1) and (63.8,169.31) .. (66.01,169.31) .. controls (68.21,169.31) and (70,171.1) .. (70,173.3) .. controls (70,175.51) and (68.21,177.29) .. (66.01,177.29) .. controls (63.8,177.29) and (62.01,175.51) .. (62.01,173.3) -- cycle ;
\draw  [fill={rgb, 255:red, 0; green, 0; blue, 0 }  ,fill opacity=1 ] (62.31,157.5) .. controls (62.31,155.4) and (64.01,153.7) .. (66.11,153.7) .. controls (68.2,153.7) and (69.9,155.4) .. (69.9,157.5) .. controls (69.9,159.59) and (68.2,161.29) .. (66.11,161.29) .. controls (64.01,161.29) and (62.31,159.59) .. (62.31,157.5) -- cycle ;
\draw  [fill={rgb, 255:red, 0; green, 0; blue, 0 }  ,fill opacity=1 ] (420.01,205.3) .. controls (420.01,203.1) and (421.8,201.31) .. (424.01,201.31) .. controls (426.21,201.31) and (428,203.1) .. (428,205.3) .. controls (428,207.51) and (426.21,209.29) .. (424.01,209.29) .. controls (421.8,209.29) and (420.01,207.51) .. (420.01,205.3) -- cycle ;
\draw  [fill={rgb, 255:red, 0; green, 0; blue, 0 }  ,fill opacity=1 ] (420.01,185.3) .. controls (420.01,183.1) and (421.8,181.31) .. (424.01,181.31) .. controls (426.21,181.31) and (428,183.1) .. (428,185.3) .. controls (428,187.51) and (426.21,189.29) .. (424.01,189.29) .. controls (421.8,189.29) and (420.01,187.51) .. (420.01,185.3) -- cycle ;
\draw  [fill={rgb, 255:red, 0; green, 0; blue, 0 }  ,fill opacity=1 ] (420.31,169.5) .. controls (420.31,167.4) and (422.01,165.7) .. (424.11,165.7) .. controls (426.2,165.7) and (427.9,167.4) .. (427.9,169.5) .. controls (427.9,171.59) and (426.2,173.29) .. (424.11,173.29) .. controls (422.01,173.29) and (420.31,171.59) .. (420.31,169.5) -- cycle ;
\draw    (388.31,140.29) -- (435.31,111.29) ;
\draw    (389.31,222.29) -- (429.31,257.29) ;

\draw (44,284) node [anchor=north west][inner sep=0.75pt]    {$p_{1}$};
\draw (25,213) node [anchor=north west][inner sep=0.75pt]    {$p_{2}$};
\draw (39,73) node [anchor=north west][inner sep=0.75pt]    {$p_{m}$};
\draw (439,85) node [anchor=north west][inner sep=0.75pt]    {$p_{m+1}$};
\draw (434,246) node [anchor=north west][inner sep=0.75pt]    {$p_{n}$};
\draw (136,90) node [anchor=north west][inner sep=0.75pt]  [font=\Large]  {$A_{1}$};
\draw (340,90) node [anchor=north west][inner sep=0.75pt]  [font=\Large]  {$A_{2}$};
\draw (255,170) node {$p$};

\end{tikzpicture}
\end{center}
\caption{The amplitude $A(p_1,\ldots,p_n)$ is built up from the lower point amplitudes $A_1(p_1,\ldots,p_m,p)$ and $A_2(p,p_{m+1},\ldots,p_n)$ by integrating over the off-shell exchanged momenta.}
\end{figure}

\noindent In momentum space, the Feynman rules imply the relation 
\begin{equation}
A(p_1,\ldots,p_n) = i\int d^4p A_1(p_1,\ldots,p_m,p)A_2(-p,p_{m+1},p_n)\Delta(-p^2)
\end{equation}
where $\Delta(-p^2)$ is the momentum space propagator, which, for scalars, depends only on $p^2$. We can perform this integral by integrating $p$ over a single mass shell $p^2 = -M^2$ and then integrating over the mass $M^2$:
\begin{equation}
\begin{split}
A(p_1,\ldots,p_n) &= i\int d^4p A_1(p_1,\ldots,p_m,p)A_2(-p,p_{m+1},p_n)\Delta(-p^2) \\
&= i\int dM^2\int d^4p\delta(p^2+M^2) A_1(p_1,\ldots,p_m,p)A_2(-p,p_{m+1},p_n)\Delta(M^2) \\
&= i\int_0^\infty dM^2\Delta(M^2)\int d^4p\delta(p^2+M^2) A_1(p_1,\ldots,p_m,p)A_2(-p,p_{m+1},\ldots,p_n) \\
 &+ i\int_{-\infty}^0 dM^2\Delta(M^2)\int d^4p\delta(p^2+M^2) A_1(p_1,\ldots,p_m,p)A_2(-p,p_{m+1},\ldots,p_n) \\
 &= i(I_+ + I_-).
\end{split}
\end{equation}
Here, $I_+$ contains the exchange of time-like momentum and $I_-$ the exchange of spacelike momentum. We deal with each in turn.\footnote{For four point functions, kinematics commonly forces either $I_-$ or $I_+$ to vanish.}

For the exchange of modes with $p^2 < 0$,
\begin{equation}
\begin{split}
I_+ &= \int_0^\infty dM^2\Delta(M^2)\int d^4p\delta(p^2+M^2)A_1(p_1,\ldots,p_m,p)A_2(-p,p_{m+1},\ldots,p_n) \\
&= \int_0^\infty dM^2M^2\Delta(M^2)\int \widetilde{d^3\hat{p}} (A_1(p_1,\ldots,p_m,M\hat{p})A_2(-M\hat{p},p_{m+1},\ldots)+(p \leftrightarrow -p)) \\
&= \int_0^\infty dM^2M^2\Delta(M^2)\int \widetilde{d^3\hat{p}}\widetilde{d^3\hat{p}'}2\hat{p}^0\delta^{(3)}(\hat{p}-\hat{p}') \\
&\times (A_1(p_1,\ldots,p_m,M\hat{p})A_2(-M\hat{p}',p_{m+1},\ldots)+A_1(p_1,\ldots,p_m,-M\hat{p})A_2(M\hat{p}',p_{m+1},\ldots)) \\
&= 2\int_0^\infty dM^2M^2\Delta(M^2)\int \widetilde{d^3\hat{p}}\widetilde{d^3\hat{p}'}\int \mu(\nu)d\nu \int dwd\bar{w}G_{1+i\nu}(\hat{p};w,\bar{w})G_{1-i\nu}(\hat{p}';w,\bar{w}) \\
&\times (A_1(p_1,\ldots,p_m,M\hat{p})A_2(-M\hat{p}',p_{m+1},\ldots)+A_1(p_1,\ldots,p_m,-M\hat{p})A_2(M\hat{p}',p_{m+1},\ldots)).
\end{split}
\end{equation}
We have used a completeness relation
\begin{equation}
\begin{split}
\int \mu(\nu)d\nu dwd\bar{w}G_{1+i\nu}&(y,z,\bar{z}|w,\bar{w})G_{1-i\nu}(y',z',\bar{z}'|w,\bar{w}) = \frac{1}{y^3}\delta(y-y')\delta^{(2)}(z-z') \\
\mu(\nu) &= \frac{\Gamma(1+i\nu)\Gamma(1-i\nu)}{4\pi^3\Gamma(i\nu)\Gamma(-i\nu)} = \frac{\nu^2}{4\pi^3}.
\end{split}
\end{equation} 
to insert two factors of the bulk-to-boundary propagator in the integral. Each integral over $\hat{p}$ now transforms the internal legs to the celestial sphere; transforming the external legs by hand gives us the relationship that 
\begin{equation}
\begin{split}
\tilde{I}_{+\Delta_1,\ldots,\Delta_n}(z_1,\ldots,z_n) = \frac{1}{2}\int_0^\infty &dM^2M^2\Delta(M^2)\int \mu(\nu)d\nu \int dw d\bar{w}  \\
&\times (\tilde{A}_{1\Delta_1,\ldots,\Delta_m,(1+i\nu,M^2)}^{\epsilon_1\ldots\epsilon_m+}(z_1,\ldots,z_m,w)\tilde{A}_{2(1-i\nu,M^2)\Delta_{m+1}\ldots\Delta_n}^{-\epsilon_{m+1}\ldots\epsilon_n}(w,z_{m+1},\ldots,z_n) \\
 &+ \tilde{A}_{1\Delta_1,\ldots,\Delta_m,(1+i\nu,M^2)}^{\epsilon_1\ldots\epsilon_m-}(z_1,\ldots,z_m,w)\tilde{A}_{2(1-i\nu,M^2)\Delta_{m+1}\ldots\Delta_n}^{+\epsilon_{m+1}\ldots\epsilon_n}(w,z_{m+1},\ldots,z_n))
\end{split}
\end{equation}
where $\tilde{I}_\pm$ is the celestial transform of $I_\pm$ and $\tilde{A} = i(\tilde{I}_+ + \tilde{I}_-)$.

We perform a similar calculation for $I_-$:
\begin{equation}
\begin{split}
I_- &= \int_{-\infty}^0 dM^2\Delta(M^2)\int d^4p\delta(p^2+M^2)A_1(p_1,\ldots,p_m,p)A_2(-p,p_{m+1},\ldots,p_n) \\
&= \int_0^\infty d\mu^2\Delta(-\mu^2)\int d^4p\delta(p^2-\mu^2)A_1(p_1,\ldots,p_m,p)A_2(-p,p_{m+1},\ldots,p_n) \\
&= \int_0^\infty d\mu^2\Delta(-\mu^2)\mu^2 \int \overline{d^3\hat{p}_+} (A_1(p_1,\ldots,p_m,\mu\hat{p}_+)A_2(-\mu\hat{p}_+,\ldots) + (\hat{p}_+ \leftrightarrow -\hat{p}_+)).
\end{split}
\end{equation}
Here we have defined $\int d^4p\delta(p^2-1)f(p) = \int \overline{d^3\hat{p}}f(\hat{p}) = \int_0^\infty \frac{d\eta dwd\bar{w}}{\eta^3}f(\eta,w,\bar{w})$, which covers half of the unit $dS_3$ slice $p^2 = 1$. Using the completeness representation derived in Appendix \ref{dscomp}
\begin{equation}
\int \mu(\nu)d\nu dwd\bar{w}G^{dS}_{1+i\nu}(\eta,z,\bar{z}|w,\bar{w})G^{dS}_{1-i\nu}(\eta',z',\bar{z}'|w,\bar{w}) = -\frac{1}{\eta^3}\delta(\eta-\eta')\delta^{(2)}(z-z')
\end{equation} 
we see that 
\begin{equation}
\begin{split}
I_- &= \int_0^\infty d\mu^2\Delta(-\mu^2)(-\mu^2) \int\overline{d^3\hat{p}_+}\overline{d^3\hat{p}_+'}(A_1(\ldots,\mu\hat{p}_+)A_2(-\mu\hat{p}_+',\ldots)+A_1(\ldots,-\mu\hat{p}_+)A_2(\mu\hat{p}_+') \\
&\times \frac{2}{\eta^3}\delta^{(3)}(\hat{p}_+-\hat{p}_+'))) \\
&= \int_0^\infty d\mu^2\Delta(-\mu^2)(-\mu^2) \int \mu(\nu)d\nu dwd\bar{w} G_{1+i\nu}^{dS}(\hat{p}_+|w,\bar{w})G_{1-i\nu}^{dS}(\hat{p}_+'|w,\bar{w}) \int \overline{d^3\hat{p}_+}\overline{d^3\hat{p}_+'}\\
&\times (A_1(\ldots,\mu\hat{p}_+)A_2(-\mu\hat{p}_+',\ldots)+A_1(\ldots,-\mu\hat{p}_+)A_2(\mu\hat{p}_+')2\hat{p}_+^0\delta^{(3)}(\hat{p}_+-\hat{p}_+'))).
\end{split}
\end{equation}
Transforming the external legs to the celestial sphere and using the bulk-to-boundary propagators to transform the internal legs gives
\begin{equation}
\begin{split}
\tilde{I}_- &= \frac{1}{2}\int_0^\infty d\mu^2\Delta(-\mu^2)(-\mu^2)\int \mu(\nu)d\nu dw d\bar{w}(	\tilde{A}^{\epsilon_1\ldots +}_{1\Delta_1,\ldots(1+i\nu,-\mu^2)}(z_1,\ldots,w)	\tilde{A}^{-\epsilon_{m+1}\ldots}_{2(1-i\nu,-\mu^2),\ldots}(w,z_{m+1},\ldots) \\
&+	\tilde{A}^{\epsilon_1\ldots -}_{1\Delta_1,\ldots(1+i\nu,-\mu^2)}(z_1,\ldots,w)	\tilde{A}^{+\epsilon_{m+1}\ldots}_{2(1-i\nu,-\mu^2),\ldots}(w,z_{m+1},\ldots)).
\end{split}
\end{equation}
We now combine these integrals to find
\begin{equation}
\label{celfeynman}
\begin{split}
\tilde{A}^{\epsilon_1\ldots\epsilon_n}_{+\Delta_1,\ldots,\Delta_n}(z_1,\ldots,z_n) = \frac{i}{2}\int_{-\infty}^\infty &dM^2M^2\Delta(M^2)\int \mu(\nu)d\nu \int dw d\bar{w}  \\
&\times (\tilde{A}_{1\Delta_1,\ldots,\Delta_m,(1+i\nu,M^2)}^{\epsilon_1\ldots\epsilon_m+}(z_1,\ldots,z_m,w)\tilde{A}_{2(1-i\nu,M^2)\Delta_{m+1}\ldots\Delta_n}^{-\epsilon_{m+1}\ldots\epsilon_n}(w,z_{m+1},\ldots,z_n) \\
 &+ \tilde{A}_{1\Delta_1,\ldots,\Delta_m,(1+i\nu,M^2)}^{\epsilon_1\ldots\epsilon_m-}(z_1,\ldots,z_m,w)\tilde{A}_{2(1-i\nu,M^2)\Delta_{m+1}\ldots\Delta_n}^{+\epsilon_{m+1}\ldots\epsilon_n}(w,z_{m+1},\ldots,z_n)).
 \end{split}
\end{equation}

This formula builds up a one-particle-reducible diagram from two lower-point diagrams. This allows us to compute all tree-level diagrams once the off-shell celestial amplitudes corresponding to the basic interactions (e.g., a massless-massless-massive three point function in a $\phi^2\Phi$ theory) are known. We can also build up any diagram from the set of off-shell celestial amplitudes corresponding to one-particle-irreducible Feynman diagrams. In section \ref{loops} we comment on the extension to loop integrals and one-particle-irreducible diagrams.
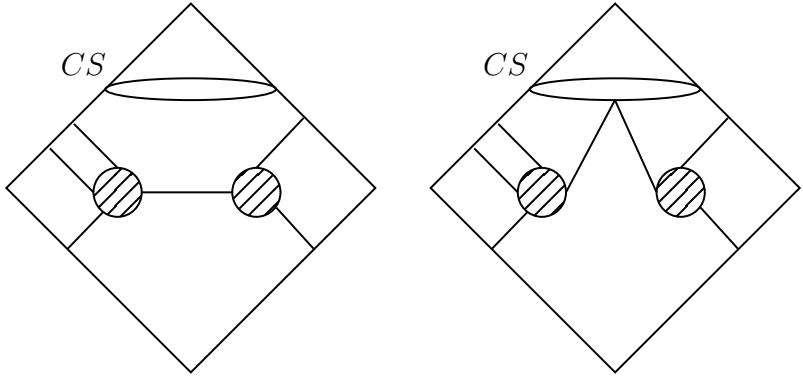
\begin{figure}[h]
\begin{center}

 
\tikzset{
pattern size/.store in=\mcSize, 
pattern size = 5pt,
pattern thickness/.store in=\mcThickness, 
pattern thickness = 0.3pt,
pattern radius/.store in=\mcRadius, 
pattern radius = 1pt}
\makeatletter
\pgfutil@ifundefined{pgf@pattern@name@_twkqkbane}{
\pgfdeclarepatternformonly[\mcThickness,\mcSize]{_twkqkbane}
{\pgfqpoint{0pt}{0pt}}
{\pgfpoint{\mcSize+\mcThickness}{\mcSize+\mcThickness}}
{\pgfpoint{\mcSize}{\mcSize}}
{
\pgfsetcolor{\tikz@pattern@color}
\pgfsetlinewidth{\mcThickness}
\pgfpathmoveto{\pgfqpoint{0pt}{0pt}}
\pgfpathlineto{\pgfpoint{\mcSize+\mcThickness}{\mcSize+\mcThickness}}
\pgfusepath{stroke}
}}
\makeatother

 
\tikzset{
pattern size/.store in=\mcSize, 
pattern size = 5pt,
pattern thickness/.store in=\mcThickness, 
pattern thickness = 0.3pt,
pattern radius/.store in=\mcRadius, 
pattern radius = 1pt}
\makeatletter
\pgfutil@ifundefined{pgf@pattern@name@_s9uc52omz}{
\pgfdeclarepatternformonly[\mcThickness,\mcSize]{_s9uc52omz}
{\pgfqpoint{0pt}{0pt}}
{\pgfpoint{\mcSize+\mcThickness}{\mcSize+\mcThickness}}
{\pgfpoint{\mcSize}{\mcSize}}
{
\pgfsetcolor{\tikz@pattern@color}
\pgfsetlinewidth{\mcThickness}
\pgfpathmoveto{\pgfqpoint{0pt}{0pt}}
\pgfpathlineto{\pgfpoint{\mcSize+\mcThickness}{\mcSize+\mcThickness}}
\pgfusepath{stroke}
}}
\makeatother

 
\tikzset{
pattern size/.store in=\mcSize, 
pattern size = 5pt,
pattern thickness/.store in=\mcThickness, 
pattern thickness = 0.3pt,
pattern radius/.store in=\mcRadius, 
pattern radius = 1pt}
\makeatletter
\pgfutil@ifundefined{pgf@pattern@name@_j57vjy324}{
\pgfdeclarepatternformonly[\mcThickness,\mcSize]{_j57vjy324}
{\pgfqpoint{0pt}{0pt}}
{\pgfpoint{\mcSize+\mcThickness}{\mcSize+\mcThickness}}
{\pgfpoint{\mcSize}{\mcSize}}
{
\pgfsetcolor{\tikz@pattern@color}
\pgfsetlinewidth{\mcThickness}
\pgfpathmoveto{\pgfqpoint{0pt}{0pt}}
\pgfpathlineto{\pgfpoint{\mcSize+\mcThickness}{\mcSize+\mcThickness}}
\pgfusepath{stroke}
}}
\makeatother

 
\tikzset{
pattern size/.store in=\mcSize, 
pattern size = 5pt,
pattern thickness/.store in=\mcThickness, 
pattern thickness = 0.3pt,
pattern radius/.store in=\mcRadius, 
pattern radius = 1pt}
\makeatletter
\pgfutil@ifundefined{pgf@pattern@name@_z26x0306f}{
\pgfdeclarepatternformonly[\mcThickness,\mcSize]{_z26x0306f}
{\pgfqpoint{0pt}{0pt}}
{\pgfpoint{\mcSize+\mcThickness}{\mcSize+\mcThickness}}
{\pgfpoint{\mcSize}{\mcSize}}
{
\pgfsetcolor{\tikz@pattern@color}
\pgfsetlinewidth{\mcThickness}
\pgfpathmoveto{\pgfqpoint{0pt}{0pt}}
\pgfpathlineto{\pgfpoint{\mcSize+\mcThickness}{\mcSize+\mcThickness}}
\pgfusepath{stroke}
}}
\makeatother
\tikzset{every picture/.style={line width=0.75pt}} 

\begin{tikzpicture}[x=0.75pt,y=0.75pt,yscale=-1,xscale=1]

\draw   (149.1,117.08) -- (242.16,210.15) -- (149.1,303.21) -- (56.04,210.15) -- cycle ;
\draw  [pattern=_twkqkbane,pattern size=6pt,pattern thickness=0.75pt,pattern radius=0pt, pattern color={rgb, 255:red, 0; green, 0; blue, 0}] (100,212.35) .. controls (100,205.53) and (105.44,200) .. (112.15,200) .. controls (118.87,200) and (124.31,205.53) .. (124.31,212.35) .. controls (124.31,219.18) and (118.87,224.71) .. (112.15,224.71) .. controls (105.44,224.71) and (100,219.18) .. (100,212.35) -- cycle ;
\draw  [pattern=_s9uc52omz,pattern size=6pt,pattern thickness=0.75pt,pattern radius=0pt, pattern color={rgb, 255:red, 0; green, 0; blue, 0}] (170,212.35) .. controls (170,205.53) and (175.44,200) .. (182.15,200) .. controls (188.87,200) and (194.31,205.53) .. (194.31,212.35) .. controls (194.31,219.18) and (188.87,224.71) .. (182.15,224.71) .. controls (175.44,224.71) and (170,219.18) .. (170,212.35) -- cycle ;
\draw    (124.31,212.35) -- (170,212.35) ;
\draw    (90.15,177.8) -- (112.15,200) ;
\draw    (78,190.15) -- (100,212.35) ;
\draw    (105,222.15) -- (87,240.8) ;
\draw    (206,174.8) -- (182.15,200) ;
\draw    (211,240.8) -- (192.15,220) ;
\draw   (106,160.3) .. controls (106,157.26) and (125.25,154.8) .. (149,154.8) .. controls (172.75,154.8) and (192,157.26) .. (192,160.3) .. controls (192,163.34) and (172.75,165.8) .. (149,165.8) .. controls (125.25,165.8) and (106,163.34) .. (106,160.3) -- cycle ;
\draw   (363.1,117.08) -- (456.16,210.15) -- (363.1,303.21) -- (270.04,210.15) -- cycle ;
\draw  [pattern=_j57vjy324,pattern size=6pt,pattern thickness=0.75pt,pattern radius=0pt, pattern color={rgb, 255:red, 0; green, 0; blue, 0}] (314,212.35) .. controls (314,205.53) and (319.44,200) .. (326.15,200) .. controls (332.87,200) and (338.31,205.53) .. (338.31,212.35) .. controls (338.31,219.18) and (332.87,224.71) .. (326.15,224.71) .. controls (319.44,224.71) and (314,219.18) .. (314,212.35) -- cycle ;
\draw  [pattern=_z26x0306f,pattern size=6pt,pattern thickness=0.75pt,pattern radius=0pt, pattern color={rgb, 255:red, 0; green, 0; blue, 0}] (384,212.35) .. controls (384,205.53) and (389.44,200) .. (396.15,200) .. controls (402.87,200) and (408.31,205.53) .. (408.31,212.35) .. controls (408.31,219.18) and (402.87,224.71) .. (396.15,224.71) .. controls (389.44,224.71) and (384,219.18) .. (384,212.35) -- cycle ;
\draw    (338.31,212.35) -- (363,165.8) ;
\draw    (304.15,177.8) -- (326.15,200) ;
\draw    (292,190.15) -- (314,212.35) ;
\draw    (319,222.15) -- (301,240.8) ;
\draw    (420,174.8) -- (396.15,200) ;
\draw    (425,240.8) -- (406.15,220) ;
\draw   (320,160.3) .. controls (320,157.26) and (339.25,154.8) .. (363,154.8) .. controls (386.75,154.8) and (406,157.26) .. (406,160.3) .. controls (406,163.34) and (386.75,165.8) .. (363,165.8) .. controls (339.25,165.8) and (320,163.34) .. (320,160.3) -- cycle ;
\draw    (363,165.8) -- (384,212.35) ;

\draw (82,140.6) node [anchor=north west][inner sep=0.75pt]    {$CS$};
\draw (295,140.6) node [anchor=north west][inner sep=0.75pt]    {$CS$};

\end{tikzpicture}
\end{center}
\caption{A Schematic diagram of how the split representation\ generates the conformal partial waves.}
\end{figure}
\section{Four Point Functions and Conformal Partial Waves}
We now specialize the result above to $n = 4$ with a single exchange. In this case, the celestial Feynman rule takes the form of a conformal partial wave expansion of the four-point function. 
\label{fpf}
\subsection{Conformal Partial Waves}
\label{fpf:cpw}

The conformal partial waves provide a complete basis for conformally-invariant four point functions that diagonalizes the quadratic Casimir acting on four points. The conformal partial waves take the form of integrals
\begin{equation}
\label{cpweq}
\Psi^{h_i,\bar{h}_i}_{\Delta,0}(z_i,\bar{z}_i) = \frac{\Gamma(\Delta)}{2\pi\Gamma(1-\Delta)}\int d^2y\langle \phi_1\phi_2\mathcal{O}_\Delta(y)\rangle_0 \langle \widetilde{\mathcal{O}_\Delta}(y)\phi_3\phi_4\rangle_0
\end{equation}
where $\langle \Oc\Oc\Oc\rangle_0$ is the appropriate covariant three-point function stripped of the structure constant $C_{ijk}$. The shadow of the operator $\Oc_\Delta$ is defined to be 
\begin{equation}
\widetilde{\Oc_\Delta}(z) = \frac{\Gamma(2-\Delta)}{\pi\Gamma(\Delta-1)}\int dzd\bar{z}\frac{1}{|z-z'|^{2(2-\Delta)}}\Oc_\Delta(z').
\end{equation}
This choice is such that the shadow transform squares to the identity \cite{do2012}. With this definition, we have that \cite{celOPEblocks}
\begin{equation}
\widetilde{\Oc_{1+i\nu}} = -2\frac{\Gamma(-i\nu)}{\Gamma(i\nu)}\Oc_{1-i\nu}.
\end{equation}
The conformal partial waves are linear combinations of the conformal blocks $k_h(z)k_{\bar{h}}(\bar{z})$ and their shadows given by 
\begin{equation}
\begin{split}
\Psi_{\Delta,J}(z_i) &= \frac{1}{2}\left(G_{h,\bar{h}}(z,\bar{z}) + \frac{K_{1-\bar{h}}}{K_h}G_{1-\bar{h},1-h}(z,\bar{z})\right) \\
K_h &= \frac{\Gamma(h\pm a)\Gamma(h\pm b)}{2\pi^2\Gamma(2h-1)\Gamma(2h)}
\end{split}
\end{equation}
where $G_{h,\bar{h}}(z,\bar{z})$ is a conformal block, $a = h_{21}$, and $b = h_{34}$ \cite{caron2017}. More generally, if the operators $\mathcal{O}$ are spinning, Equation \ref{cpweq} will generate conformal partial waves labeled by a weight $\Delta$ and an integral spin $J$, the spin of $\mathcal{O}$. These partial waves provide a complete, orthogonal,  $\delta$-function normalizable basis for conformally covariant four-point functions and are commonly used to extract conformal block decompositions from four-point functions.

Any conformally-covariant four-point function can be expanded in conformal partial waves
\begin{equation}
	G(z_i,\bar{z}_i) = 1_{12}1_{34} + \sum_{J=0}^\infty \int_{1-i\infty}^{1+i\infty}\frac{d\Delta}{2\pi i}c(\Delta,J)\Psi_{\Delta,J}(z_i,\bar{z}_i).
\end{equation}
If we gauge fix so that $G(z_i,\bar{z}_i) = I_{12-34}(z_i,\bar{z}_i)\mathcal{G}(z,\bar{z})$ and $\Psi_{\Delta,J} = I_{12-34}(z_i,\bar{z}_i)\Psi_{\Delta,J}(z,\bar{z})$ where $I_{12-34}(z_i,\bar{z}_i)$ is the standard conformally covariant prefactor defined in Equation \ref{fptca} and $z$ is the cross ratio, the conformal wave coefficients can be found using the Euclidean inversion formula
\begin{equation}
c(\Delta,J) = \frac{K_h}{2\pi K_{1-\bar{h}}}\int d^2z\frac{(1-z)^{a+b}(1-\bar{z})^{\bar{a}+\bar{b}}}{(z\bar{z})^2}\Psi_{\Delta,J}(z,\bar{z})\mathcal{G}(z,\bar{z})
\end{equation}
where the integral runs over the complex plane $\bar{z} = z^*$ \cite{caron2017}.\footnote{In some cases non-normalizable parts of the four-point function must be subtracted out by hand before performing the conformal partial wave decomposition.} Because the conformal partial waves are linear combinations of conformal blocks, evaluating the integrals as a sum over residues of the coefficients of the conformal blocks gives the familiar conformal block decomposition.
\subsection{Celestial Feynman Expansion of Four-Point Functions}
\label{fpf:cpwisf}
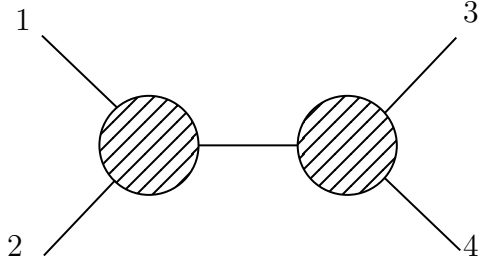
\begin{figure}[h!]
\begin{center}

 
\tikzset{
pattern size/.store in=\mcSize, 
pattern size = 5pt,
pattern thickness/.store in=\mcThickness, 
pattern thickness = 0.3pt,
pattern radius/.store in=\mcRadius, 
pattern radius = 1pt}
\makeatletter
\pgfutil@ifundefined{pgf@pattern@name@_bqil829u1}{
\pgfdeclarepatternformonly[\mcThickness,\mcSize]{_bqil829u1}
{\pgfqpoint{0pt}{0pt}}
{\pgfpoint{\mcSize+\mcThickness}{\mcSize+\mcThickness}}
{\pgfpoint{\mcSize}{\mcSize}}
{
\pgfsetcolor{\tikz@pattern@color}
\pgfsetlinewidth{\mcThickness}
\pgfpathmoveto{\pgfqpoint{0pt}{0pt}}
\pgfpathlineto{\pgfpoint{\mcSize+\mcThickness}{\mcSize+\mcThickness}}
\pgfusepath{stroke}
}}
\makeatother

 
\tikzset{
pattern size/.store in=\mcSize, 
pattern size = 5pt,
pattern thickness/.store in=\mcThickness, 
pattern thickness = 0.3pt,
pattern radius/.store in=\mcRadius, 
pattern radius = 1pt}
\makeatletter
\pgfutil@ifundefined{pgf@pattern@name@_omuk34zvj}{
\pgfdeclarepatternformonly[\mcThickness,\mcSize]{_omuk34zvj}
{\pgfqpoint{0pt}{0pt}}
{\pgfpoint{\mcSize+\mcThickness}{\mcSize+\mcThickness}}
{\pgfpoint{\mcSize}{\mcSize}}
{
\pgfsetcolor{\tikz@pattern@color}
\pgfsetlinewidth{\mcThickness}
\pgfpathmoveto{\pgfqpoint{0pt}{0pt}}
\pgfpathlineto{\pgfpoint{\mcSize+\mcThickness}{\mcSize+\mcThickness}}
\pgfusepath{stroke}
}}
\makeatother
\tikzset{every picture/.style={line width=0.75pt}} 

\begin{tikzpicture}[x=0.75pt,y=0.75pt,yscale=-1,xscale=1]

\draw  [pattern=_bqil829u1,pattern size=6pt,pattern thickness=0.75pt,pattern radius=0pt, pattern color={rgb, 255:red, 0; green, 0; blue, 0}] (140,175) .. controls (140,161.19) and (151.19,150) .. (165,150) .. controls (178.81,150) and (190,161.19) .. (190,175) .. controls (190,188.81) and (178.81,200) .. (165,200) .. controls (151.19,200) and (140,188.81) .. (140,175) -- cycle ;
\draw  [pattern=_omuk34zvj,pattern size=6pt,pattern thickness=0.75pt,pattern radius=0pt, pattern color={rgb, 255:red, 0; green, 0; blue, 0}] (240,175) .. controls (240,161.19) and (251.19,150) .. (265,150) .. controls (278.81,150) and (290,161.19) .. (290,175) .. controls (290,188.81) and (278.81,200) .. (265,200) .. controls (251.19,200) and (240,188.81) .. (240,175) -- cycle ;
\draw    (190,175) -- (240,175) ;
\draw    (111,119.6) -- (149,156) ;
\draw    (284,191.6) -- (322,228) ;
\draw    (284,158.6) -- (320,120.6) ;
\draw    (112,230.6) -- (148,192.6) ;

\draw (96,105) node [anchor=north west][inner sep=0.75pt]    {$1$};
\draw (92,219) node [anchor=north west][inner sep=0.75pt]    {$2$};
\draw (322,101) node [anchor=north west][inner sep=0.75pt]    {$3$};
\draw (322,219) node [anchor=north west][inner sep=0.75pt]    {$4$};

\end{tikzpicture}
\end{center}
\caption{A four point function with exchange in the 12-34 channel.}\label{fig:fourpoint}
\end{figure}

If we specialize to four-point amplitudes, the integral over $w,\bar{w}$ of the product of three-point structures reproduces a conformal partial wave, showing that the CPW coefficients are integrals over products of off-shell three-point structures. Consider the amplitude $A$ corresponding to the schematic Feynman diagram in Figure \ref{fig:fourpoint}.

\noindent The three-point structure with $M^2 \ne 0$ takes the form 
\begin{equation}
    	\tilde{A}^{\epsilon_1\epsilon_2\epsilon_3}_{\Delta_1,\Delta_2,(\Delta_3,M^2)} = \frac{C^{\epsilon_1\epsilon_2\epsilon_3}_{\Delta_1,\Delta_2,(\Delta_3,M^2)}}{|z_{12}|^{\Delta_1+\Delta_2-\Delta_3}|z_{23}|^{\Delta_2+\Delta_3-\Delta_1}|z_{13}|^{\Delta_1+\Delta_3-\Delta_2}}
\end{equation}
since at least one leg is massive \cite{poinconst}. Applying Equation \ref{celfeynman} gives
	\begin{equation}
	\begin{split}
		\tilde{A}^{\epsilon_1\epsilon_2\epsilon_3\epsilon_4}_{\Delta_1,\Delta_2,\Delta_3,\Delta_4}(z_1,z_2,z_3,z_4) &= \frac{i}{2}\int_{-\infty}^\infty M^2\Delta(M^2) dM^2\int \mu(\nu)d\nu \int d^2w \\
		&\times (	\tilde{A}^{\epsilon_1\epsilon_2+}_{\Delta_1,\Delta_2,(1+i\nu,M^2)}(z_1,z_2,w)	\tilde{A}^{-\epsilon_3\epsilon_4}_{(1-i\nu,M^2),\Delta_3,\Delta_4}(w,z_3,z_4) \\
		&+	\tilde{A}^{\epsilon_1\epsilon_2-}_{\Delta_1,\Delta_2,(1+i\nu,M^2)}(z_1,z_2,w)	\tilde{A}^{+\epsilon_3\epsilon_4}_{(1-i\nu,M^2),\Delta_3,\Delta_4}(w,z_3,z_4)) \\
		&=-\frac{\pi}{2}\int M^2\Delta(M^2) dM^2 \int \frac{\mu(\nu)d\nu}{\nu} (C^{\epsilon_1\epsilon_2+}_{12(1+i\nu,M^2)}C^{-\epsilon_3\epsilon_4}_{(1-i\nu,M^2)34} \\
		&+C^{\epsilon_1\epsilon_2-}_{12(1+i\nu,M^2)}C^{+\epsilon_3\epsilon_4}_{(1-i\nu,M^2)34}) \\
		&\times \Psi_{1+i\lambda}(z_1,z_2,z_3,z_4)
		\end{split}
	\end{equation}
where $\Psi_{1+i\lambda}(z_i)$ is a conformal partial wave. Recognizing the inverse two point function of massive scalars $D^{(1+i\nu,M^2),(1+i\nu,M^2)} = \frac{i\nu M^2}{2\pi}$, this takes the form 
	\begin{equation}
	\begin{split}
	    \tilde{A}^{\epsilon_1\epsilon_2\epsilon_3\epsilon_4}_{\Delta_1,\Delta_2,\Delta_3,\Delta_4}(z_1,z_2,z_3,z_4) &= \int_{-\infty}^\infty \frac{i\Delta(M^2)dM^2}{2}\int \frac{d\nu}{2\pi} D^{(1+i\nu,M^2),(1+i\nu,M^2)} \\
	    &\times (C^{\epsilon_1\epsilon_2+}_{12(1+i\nu,M^2)}C^{-\epsilon_3\epsilon_4}_{(1-i\nu,M^2)34} +C^{\epsilon_1\epsilon_2-}_{12(1+i\nu,M^2)}C^{+\epsilon_3\epsilon_4}_{(1-i\nu,M^2)34}) \\
		&\times \Psi_{1+i\lambda}(z_1,z_2,z_3,z_4).
	\end{split}
	\end{equation}
We see that the coefficient of the conformal partial wave is an integral over a product of three-point functions where the internal line is off-shell times a propagator that depends on the square of the exchanged off-shell momentum. The inverse two-point function of off-shell massive particles, which goes as $\nu$, enters through a product of the measure $\mu(\nu)$ and a conversion factor relating the structure $\int dwd\bar{w}\langle \Oc\Oc\Oc_{1+i\nu}\rangle_0\langle \Oc_{1-\nu}\Oc\Oc\rangle_0$ to the conformal partial wave. 

Interestingly, we see that the three-point structures in celestial conformal field theory arise at the level of the conformal partial wave decomposition. This partial wave decomposition also gives only a single channel for the scattering amplitudes. Other Feynman diagrams that contribute to the amplitude will contribute terms that are integrals over conformal partial waves in different channels which can then be related by crossing to form a partial wave expansion in a single channel.

This also makes it clear why we should expect to see an expansion in terms of the familiar conformal partial waves $\Psi_{1+i\lambda}$ even for theories with only massless particles, for which the on-shell three point functions vanish in (3,1) signature and are distributional in (2,2) signature. Because the exchanged momentum generically has $p^2 \ne 0$ and three-point functions of scalars with at least one massive leg are the familiar conformally covariant non-singular three-point structure, there exists a meaningful decomposition into the conformal partial waves \cite{poinconst}. Off-shell three point functions $C_{\Delta_1,\Delta_2,(\Delta_3,0)}$ contribute to the expansion non-trivially only when the physical mass of the exchanged particle vanishes, in which case the propagator develops a $\delta$-function like imaginary part at $p^2 = 0$. The contribution from massless exchange in this case vanishes in (3,1)-signature Minkowski space due to kinematic constraints on amplitudes with three external massless legs. This contribution can be studied in (2,2)-signature Klein space.

In conformal field theories, we expect conformal block expansions in the $12-34$ channel to have coefficients that factor into a function of operators 1, 2, and the exchange times a function of operators 3, 4, and the exchange. However, in celestial CFTs, four-point functions of massless external particles generically have a prefactor that depends only on the sum of the external dimensions and encodes effective field theory expansions of the amplitude in its poles and residues \cite{UVIR}. This prefactor prevents the conformal block expansion from factorizing as we expect in such a CFT \cite{CPW}. Here, we see the origin of the factorization problems in these conformal partial wave calculations: the integral over $M^2 = -p^2$. Because the two-point function must vanish between operators of different masses, this integral cannot factor as we would expect in a conformal field theory unless the spectrum is expanded to include off-shell exchanged states.  Because the integrand over $M^2$ develops an imaginary part only for $M^2 = m^2$ for massive particles, applying the optical theorem localizes the integral to on-shell exchange and yields a four-point function with a factorizable conformal partial wave expansion of the imaginary part of the four-point function \cite{shaolam}.

\section{A Simple Four-Point Scalar CPW Decomposition}
\label{scalarfpf}

As an example, consider a tree level $s$-channel celestial amplitude of four external massless scalars mediated by massive exchange:
\begin{figure}[h!]
    \centering

\tikzset{every picture/.style={line width=0.75pt}} 

\begin{tikzpicture}[x=0.75pt,y=0.75pt,yscale=-1,xscale=1]

\draw    (520,150) -- (520,210) ;
\draw [shift={(520,210)}, rotate = 90] [color={rgb, 255:red, 0; green, 0; blue, 0 }  ][fill={rgb, 255:red, 0; green, 0; blue, 0 }  ][line width=0.75]      (0, 0) circle [x radius= 3.35, y radius= 3.35]   ;
\draw [shift={(520,150)}, rotate = 90] [color={rgb, 255:red, 0; green, 0; blue, 0 }  ][fill={rgb, 255:red, 0; green, 0; blue, 0 }  ][line width=0.75]      (0, 0) circle [x radius= 3.35, y radius= 3.35]   ;
\draw    (480,240) -- (520,210) ;
\draw    (520,150) -- (560,120) ;
\draw    (480,120) -- (520,150) ;
\draw    (520,210) -- (560,240) ;

\draw (469,239) node [anchor=north west][inner sep=0.75pt]    {$1$};
\draw (560,240) node [anchor=north west][inner sep=0.75pt]    {$2$};
\draw (465,99) node [anchor=north west][inner sep=0.75pt]    {$3$};
\draw (559,99) node [anchor=north west][inner sep=0.75pt]    {$4$};

\end{tikzpicture}
    \caption{An $s$ channel massive exchange diagram for a four massless scalar amplitude mediated by massive exchange.}
    \label{fig:examp}
\end{figure}
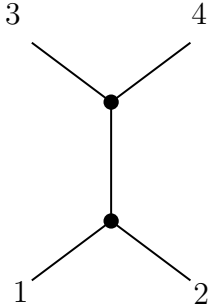

\noindent This corresponds to the momentum space amplitude 
\begin{equation}
    A(p_1,p_2,p_3,p_4) = \frac{g^2}{-s^2+m^2}\delta^{(4)}(p_1,p_2,p_3,p_4)
\end{equation}
and the celestial amplitude \cite{volcpw}
\begin{equation}
\label{fptca}
\begin{split}
    \tilde{A}_{\Delta_1,\Delta_2,\Delta_3,\Delta_4}^{++--}(z_1,z_2,z_3,z_4) &= I_{12-34}(z_i,\bz_i) \times \frac{g^2(im/2)^{\beta-2}}{\sin(\pi\beta/2)}z^2(1-z)^{h_{12}-h_{34}}\delta(z-\bar{z}) \\
    I_{12-34}(z_i,\bz_i) &= \frac{\left(\frac{z_{24}}{z_{14}}\right)^{h_{12}}\left(\frac{z_{14}}{z_{13}}\right)^{h_{34}}}{z_{12}^{h_1+h_2}z_{34}^{h_3+h_4}}\frac{\left(\frac{\bz_{24}}{\bz_{14}}\right)^{\bz_{12}}\left(\frac{\bz_{14}}{\bz_{13}}\right)^{\bz_{34}}}{\bz_{12}^{h_1+h_2}\bz_{34}^{h_3+h_4}}\\
    z &= \frac{z_{12}z_{34}}{z_{13}z_{24}},\ \beta = \sum_i\Delta_i-4
\end{split}
\end{equation}
The off-shell massless-massless-massive three-point function for $\Delta_j = 1+i\lambda_j$ is 
\begin{equation}
C^{++-}_{ij(1+i\lambda,M^2)} = C^{--+}_{ij(1+i\lambda,M^2)}= \frac{g}{2M^2}\left(\frac{M}{2}\right)^{i\lambda_i+i\lambda_j}B\left(\frac{1+i\lambda+i\lambda_{ij}}{2},\frac{1+i\lambda-i\lambda_{ij}}{2}\right)
\end{equation}
for $M^2 > 0$. All other ingoing/outgoing configurations and $C_{ij(1+i\lambda,-\mu^2)}$ vanish \cite{RaclariuReview}. 
This leads to the conformal partial wave expansion
\begin{equation}
\begin{split}
\tilde{A}^{++--}_{\Delta_1\Delta_2\Delta_3\Delta_4}(z_i) &= \int_0^\infty \frac{i\Delta(-M^2)dM^2}{2} \int \frac{d\nu}{2\pi} D^{(1+i\nu,M^2),(1+i\nu,M^2)} C_{12(1+i\nu,M^2)}^{++-}C_{(1-i\nu,M^2)34}^{-++}\Psi_{1+i\lambda}(z_i)) \\
&= \int_0^\infty \frac{dM^2}{2(-M^2+m^2-i\epsilon)} \int \frac{d\nu}{2\pi}D^{(1+i\nu,M^2),(1+i\nu,M^2)} C_{12(1+i\nu,M^2)}^{++-}C_{(1-i\nu,M^2)34}^{-++}\Psi_{1+i\lambda}(z_i).
\end{split}
\end{equation}
Because $A^{++-}_{12(\Delta,-\mu^2)} = A^{--+}_{(\Delta,-\mu^2)34} = 0$, the integral in $M^2$ is one sided.

\section{Loops}
\label{loops}
In momentum space, loop integrals can also be computed by integrating over multiple exchanged momenta. In momentum space, with $p_i$ being the external momenta and $p_j'$ the internal momenta, the Feynman rules take the form 
\begin{equation}
    A(p_i) = \frac{i}{S}\int \prod_{j=1}^N p_j' A_1(p_1,\ldots,p_m,p_1',\ldots,p_N')\prod_{j=1}^N \Delta(-{p_j'}^2) A_2(-p_1',\ldots,p_N',p_{m+1},\ldots,p_n)
\end{equation} 
corresponding to a diagram with $N$ internal legs. Here $S$ is accounts for any symmetry factors not containted in $A_1$ and $A_2$.
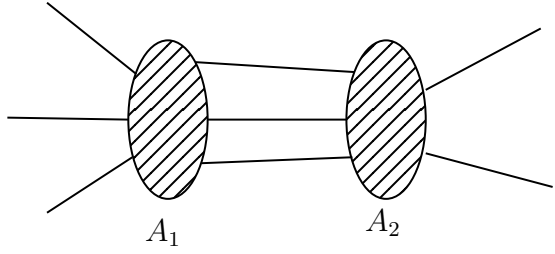
\begin{figure}[h]
\begin{center}

 
\tikzset{
pattern size/.store in=\mcSize, 
pattern size = 5pt,
pattern thickness/.store in=\mcThickness, 
pattern thickness = 0.3pt,
pattern radius/.store in=\mcRadius, 
pattern radius = 1pt}
\makeatletter
\pgfutil@ifundefined{pgf@pattern@name@_1z01afr7c}{
\pgfdeclarepatternformonly[\mcThickness,\mcSize]{_1z01afr7c}
{\pgfqpoint{0pt}{0pt}}
{\pgfpoint{\mcSize+\mcThickness}{\mcSize+\mcThickness}}
{\pgfpoint{\mcSize}{\mcSize}}
{
\pgfsetcolor{\tikz@pattern@color}
\pgfsetlinewidth{\mcThickness}
\pgfpathmoveto{\pgfqpoint{0pt}{0pt}}
\pgfpathlineto{\pgfpoint{\mcSize+\mcThickness}{\mcSize+\mcThickness}}
\pgfusepath{stroke}
}}
\makeatother

 
\tikzset{
pattern size/.store in=\mcSize, 
pattern size = 5pt,
pattern thickness/.store in=\mcThickness, 
pattern thickness = 0.3pt,
pattern radius/.store in=\mcRadius, 
pattern radius = 1pt}
\makeatletter
\pgfutil@ifundefined{pgf@pattern@name@_ajff2d3hq}{
\pgfdeclarepatternformonly[\mcThickness,\mcSize]{_ajff2d3hq}
{\pgfqpoint{0pt}{0pt}}
{\pgfpoint{\mcSize+\mcThickness}{\mcSize+\mcThickness}}
{\pgfpoint{\mcSize}{\mcSize}}
{
\pgfsetcolor{\tikz@pattern@color}
\pgfsetlinewidth{\mcThickness}
\pgfpathmoveto{\pgfqpoint{0pt}{0pt}}
\pgfpathlineto{\pgfpoint{\mcSize+\mcThickness}{\mcSize+\mcThickness}}
\pgfusepath{stroke}
}}
\makeatother
\tikzset{every picture/.style={line width=0.75pt}} 

\begin{tikzpicture}[x=0.75pt,y=0.75pt,yscale=-1,xscale=1]

\draw  [pattern=_1z01afr7c,pattern size=6pt,pattern thickness=0.75pt,pattern radius=0pt, pattern color={rgb, 255:red, 0; green, 0; blue, 0}] (260,70) .. controls (271.05,70) and (280,87.91) .. (280,110) .. controls (280,132.09) and (271.05,150) .. (260,150) .. controls (248.95,150) and (240,132.09) .. (240,110) .. controls (240,87.91) and (248.95,70) .. (260,70) -- cycle ;
\draw  [pattern=_ajff2d3hq,pattern size=6pt,pattern thickness=0.75pt,pattern radius=0pt, pattern color={rgb, 255:red, 0; green, 0; blue, 0}] (370,70) .. controls (381.05,70) and (390,87.91) .. (390,110) .. controls (390,132.09) and (381.05,150) .. (370,150) .. controls (358.95,150) and (350,132.09) .. (350,110) .. controls (350,87.91) and (358.95,70) .. (370,70) -- cycle ;
\draw    (274,81) -- (354,86) ;
\draw    (280,110) -- (350,110) ;
\draw    (277,132) -- (353,129) ;
\draw    (199,51) -- (244,87) ;
\draw    (179,109) -- (240,110) ;
\draw    (199,157) -- (242,129) ;
\draw    (390,127) -- (454,144) ;
\draw    (390,95) -- (448,64) ;

\draw (247,159) node [anchor=north west][inner sep=0.75pt]    {$A_{1}$};
\draw (358,154) node [anchor=north west][inner sep=0.75pt]    {$A_{2}$};

\end{tikzpicture}
\caption{A Feynman diagram with $N=3$ internal legs and 5 external legs.\label{fig:loop}}
\end{center}
\end{figure}

\noindent Transforming each external leg to the celestial sphere and applying the derivation above to each internal leg gives the celestial loop integral
\begin{equation}
\label{celloop}
  \begin{split}
        \tilde{A}_{\Delta_1\cdots\Delta_n}(z_1,\ldots,z_n) &=   \frac{i}{S2^N}\int \prod_{j=1}^N M_j^2dM_j^2\Delta_j(M_j^2) \mu(\nu_j)d\nu_jdw_jd\bar{w}_j \\
        &\times \sum_{\{\epsilon_j' \in \{\pm\}^N\}} \tilde{A}^{\epsilon_1\cdots\epsilon_m\epsilon_1'\cdots\epsilon_N'}_{1\Delta_1\cdots \Delta_m (1+i\nu_1,M_1^2)\cdots(1+i\nu_N,M_N^2)}(z_1,\ldots,z_m,w_1,\ldots,w_N) \\
        &\times \tilde{A}^{(-\epsilon_1')\cdots(-\epsilon_N')\epsilon_{m+1}\cdots\epsilon_m}_{2(1-i\nu_1,M_1^2)\cdots(1-i\nu_N,M_N^2)\Delta_{m+1}\cdots\Delta_n}(w_1,\ldots,w_N,z_{m+1},\ldots,z_n).
    \end{split}
\end{equation}
In momentum space, loop integrals often diverge and require regularization. We leave an understanding of how to regularize the celestial loop integrals in Equation \ref{celloop} to future work.
\section*{Acknowledgements}

I am grateful to Alex Atanasov, Adam Ball, Ana-Maria Raclariu, and Andrew Strominger  for useful conversations, and to Ana-Maria Raclariu and Andrew Strominger for comments on an early draft of this work. This material is based upon work supported by the National Science Foundation Graduate Research Fellowship Program under Grant No. DGE1745303. Any opinions, findings, and conclusions or recommendations expressed in this material are those of the author and do not necessarily reflect the views of the National Science Foundation.
\appendix

\section{Analytic Continuation from $M^2 > 0$}
\label{ancon}
For $M^2 > 0$ the off-shell conformal primary wavefunction is defined by 
\begin{equation}
\phi^\pm_{(\Delta,M^2)}(X|z,\bar{z}) = \int \frac{dydwd\bar{w}}{y^3}G_{\Delta}(y,w,\bar{w}|z,\bar{z})e^{\pm im\hat{p} \cdot X}.
\end{equation}
This integral can be evaluated by analytic continuation from $m \in -i\mathbb{R}_+$, where the integral takes the form 
\begin{equation}
  \phi^\pm_{(\Delta,M^2)}(X|z,\bar{z}) = \int \frac{dydwd\bar{w}}{y^3}G_{\Delta}(y,w,\bar{w}|z,\bar{z})e^{\pm\mu\hat{p} \cdot X}  
\end{equation}
 and is convergent for $X^2 < 0$ \cite{fsa2017}. We define off-shell celestial amplitudes with spacelike legs by an integral over oscillatory solutions of spacelike real momentum and define them so that they agree with the wavefunction found by analytically continuing in $M$. Analytically continuing $M \to -i\mu \in -i\mathbb{R}_+$ gives 
\begin{equation}
\begin{split}
\phi^\pm_{(\Delta,-\mu^2)}(X|z,\bar{z}) &=\int_0^\infty \frac{dy}{y^3}\int dwd\bar{w} G_{\Delta}(y,w,\bar{w}|z,\bar{z})e^{\pm i\mu(-i\hat{p}\cdot X)}.
\end{split}
\end{equation}
Defining $\eta = iy$, this gives
\begin{equation}
\begin{split}
\phi^\pm_{(\Delta,-\mu^2)}(X|z,\bar{z}) &=-\int_0^{i\infty} \frac{d\eta}{\eta^3}\int dwd\bar{w} G_{\Delta}(-i\eta,w,\bar{w}|z,\bar{z})e^{\pm i\mu(\hat{p}_+(\eta,w,\bar{w})\cdot X)} \\
&= \int_0^\infty\frac{d\eta}{\eta^3}\int dwd\bar{w} \left(-G_\Delta(-i\eta,w,\bar{w}|z,\bar{z})\right)e^{\pm i\mu\hat{p}_+ \cdot X}
\end{split}
\end{equation}
where in the last step we have deformed the contour from the positive imaginary axis to the positive real axis. Therefore, we define
\begin{equation}
G^{dS}_\Delta(\eta,w,\bar{w}|z,\bar{z}) = -G_\Delta(-i\eta,w,\bar{w}|z,\bar{z}) = -\left(\frac{-i\eta}{-\eta^2+|z-w|^2}\right)^\Delta
\end{equation}
so that the off-shell conformal primary wavefunctions are analytic in $M^2$.  

\section{Completeness Relations for $G^{dS}_\Delta$}
\label{dscomp}
In this section we find the completeness relation for $G^{dS}_\Delta$. Letting $F(\hat{p}_1,\hat{p}_2)$ be any function of two unit timelike momenta, the $H_3$ bulk-to-boundary propagators satisfy 
\begin{equation}
\int_0^\infty \frac{dy_1}{y_1^3}\int dw_1d\bar{w}_1 \int \mu(\nu)d\nu dzd\bar{z} G_{1+i\nu}(\hat{p}_1|z,\bar{z})G_{1-i\nu}(\hat{p}_2|z,\bar{z})F(\hat{p}_1,\hat{p}_2) = F(\hat{p}_2,\hat{p}_2).
\end{equation}
Redefining the variable of integration to $\eta = iy$ we have that 
\begin{equation}
\begin{split}
    \int_0^{i\infty} \frac{d\eta_1}{\eta_1^3}\int dw_1d\bar{w}_1\int\mu(\nu)d\nu dzd\bar{z} (-&G_{1+i\nu}(-i\eta_1,w_1,\bar{w}_1|z,\bar{z})G_{1-i\nu}(-i\eta_2,w_2,\bar{w}_2|z,\bar{z}) \\
     &= F(-i\eta_2,w_2,\bar{w}_2,-i\eta_2,w_2,\bar{w}_2)
\end{split}
\end{equation}
where we have defined $y_2 = -i\eta_2$. Analytically continuing and rotating the contour of integration gives us the relation 
\begin{equation}
    \begin{split}
        \int_0^\infty \frac{d\eta_1}{\eta_1^3}\int dw_1d\bar{w}_1\int\mu(\nu)d\nu dzd\bar{z} G^{dS}_{1+i\nu}(\eta_1,w_1,\bar{w}_1|z,\bar{z})G^{dS}_{1-i\nu}(\eta_2,w_2,\bar{w}_2|z,\bar{z})&F(\hat{p}_1^+,\hat{p}_2^+) \\
         &= -F(\hat{p}_2^+,\hat{p}_2^+).
    \end{split}
\end{equation}
Since $F$ was an arbitrary test function, this implies the completeness relation 
\begin{equation}
    \int \mu(\nu)d\nu dzd\bar{z} G_{1+i\nu}^{dS}(\eta_1,w_1,\bar{w}_1|z,\bar{z})G_{1-i\nu}^{dS}(\eta_2,w_2,\bar{w}_2|z,\bar{z}) = -\eta_1^3\delta(\eta_1-\eta_2)\delta^{(2)}(w_1-w_2).
\end{equation}

\bibliographystyle{utphys}
\bibliography{bib}

\end{document}